\documentclass[prb,preprint,showpacs,preprintnumbers,amsmath,amssymb]{revtex4}

\usepackage{graphicx}
\usepackage{dcolumn}
\usepackage{bm}

\begin{document}


\title{$N$-representability of the Jastrow wave function pair density 
of the lowest-order}

\author{Katsuhiko Higuchi}
\email{khiguchi@hiroshima-u.ac.jp}
\affiliation{Graduate School of Advanced Sciences of Matter, 
Hiroshima University, Higashi-Hiroshima 739-8527, Japan}
\author{Masahiko Higuchi}%
\email{higuchi@shinshu-u.ac.jp}
\affiliation{Department of Physics, Faculty of Science, 
Shinshu University, Matsumoto 390-8621, Japan}

\date{\today}

\begin{abstract}
We have recently proposed a density functional scheme for calculating the 
ground-state pair density (PD) within the Jastrow wave function PDs of the 
lowest-order (LO-Jastrow PDs) [M. Higuchi and K. Higuchi, Phys. Rev. A 
\textbf{75}, 042510 (2007)]. However, there remained an arguable problem on 
the $N$-representability of the LO-Jastrow PD. In this paper, the sufficient 
conditions for the $N$-representability of the LO-Jastrow PD are derived. 
These conditions are used as the constraints on the correlation function of 
the Jastrow wave function. A concrete procedure to search the suitable 
correlation function is also presented. 
\end{abstract}


\maketitle

\section{INTRODUCTION}
The pair density (PD) functional theory has been expected to be one of the 
promising schemes beyond the density functional theory.\cite{1} Recently, we 
have proposed the PD functional theory that yields the best PD within the 
set of the Jastrow wave function PDs of the lowest-order 
(LO-Jastrow PDs).\cite{2} 
The search region for the ground-state PD is substantially extended as 
compared with the previous theory.\cite{3,4}
On the other hand, however, there 
remains a significant problem related to the $N$-representability of the 
LO-Jastrow PDs.\cite{2} 

Let us revisit the problem here. We shall consider an $N_{0} $-electron 
system. The Jastrow wave function is given by
\begin{equation}
\label{eq1}
\Psi _{J} \left({x_{1}, \cdot \cdot \cdot, x_{N_{0}} } \right) = 
\frac{1}{\sqrt {A_{N_{0} } } }\prod\limits_{1 \le i < j \le N_{0} } 
{f\left( {\left| {{\rm {\bf r}}_{i} - {\rm {\bf r}}_{j} } \right|} \right)} 
\Phi_{SSD} \left( {x_{1}, \cdot \cdot \cdot, x_{N_{0} } } \right),
\end{equation}
where $x_{i} $ denotes the coordinates including the spatial coordinate 
${\rm {\bf r}}_{i} $ and spin coordinate $\eta _{i} $, and where $A_{N_{0} } 
$, $f\left( {\left| {{\rm {\bf r}}_{i} - {\rm {\bf r}}_{j} } \right|} 
\right)$ and $\Phi _{SSD} \left( {x_{1} ,\, \cdot \cdot \cdot ,\,x_{N_{0} } 
} \right)$ are the normalization constant, correlation function and single 
Slater determinant (SSD), respectively. The LO-Jastrow PD is given by 
\cite{2,5,6} 
\begin{equation}
\label{eq2}
\gamma_{LO}^{(2)} \left( {{\rm {\bf r{r}'}};{\rm {\bf r{r}'}}} 
\right) = \left| {f\left( {\left| {{\rm {\bf r}}_{i} - {\rm {\bf r}}_{j} } 
\right|} \right)} \right|^{2}
\gamma_{SSD}^{(2)} \left( {{\rm {\bf r{r}'}};{\rm {\bf r{r}'}}} \right)
_{N = N_{0} } ,
\end{equation}
where $\gamma_{SSD}^{(2)} \left( {{\rm {\bf r{r}'}};{\rm {\bf 
r{r}'}}} \right)_{N = N_{0} }$ 
is the PD calculated from the SSD.  In the preceding paper,\cite{2} 
we have confirmed that Eq. (\ref{eq2}) meets four kinds of 
necessary conditions for the $N$-representability of the PD, and may become 
"approximately $N$-representable".\cite{7,8,9,10,11,12}  
However, the possibility of it 
being $N$-representable has not been discussed.\cite{2} 
This is an arguable problem that is concerned with whether the reproduced PD 
is physically reasonable or not. 

The aim of this paper is to discuss the $N$-representability 
of Eq.(\ref{eq2}) and to show the way to search the suitable correlation 
function 
$f\left( {\left| {{\rm {\bf r}}_{i} - {\rm {\bf r}}_{j} } \right|} \right)$. 
The organization of this paper is as follows.  
For the convenience of the subsequent discussions, we first examine 
the properties of the LO-Jastrow PD in Sec. II.  
The sufficient conditions for the $N$-representability of 
Eq. (\ref{eq2}), which are imposed on the correlation function, 
will be derived recursively in Sec. III.  
Concrete steps for searching the correlation function that meets these 
conditions are discussed in Sec. IV.  
Finally, concluding remarks are given in Sec. V. 

\section{Properties of the LO-Jastrow PD}
In this section, we shall discuss the properties of the LO-Jastrow PD. 
To this aim, the properties of PDs that are calculated from SSDs are 
investigated. The cofactor expansion of 
$\Phi _{SSD} (x_{1} , \cdot \cdot \cdot ,x_{N_{0} } )$ 
along the $N_{0} $th row leads to 
\begin{eqnarray}
\label{eq3}
\Phi_{SSD} (x_{1}, \cdot \cdot \cdot, x_{N_{0} }) = 
\frac{1}{\sqrt{N_{0}}} &\{& 
(-1)^{N_{0} + 1}\phi_{\lambda }(x_{N_{0}})
\Phi_{SSD}^1 (x_{1}, \cdot \cdot \cdot, x_{N_{0} - 1}) \nonumber \\ 
&+& (-1)^{N_{0}+2}\phi_{\mu} (x_{N_{0}})
\Phi_{SSD}^2 (x_{1}, \cdot \cdot \cdot, x_{N_{0}-1} ) \nonumber \\
&+& \cdot \cdot \cdot \\ 
&+& (-1)^{2N_{0}}\phi_{\xi} (x_{N_{0} })
\Phi_{SSD}^{N_{0}} (x_{1}, \cdot \cdot \cdot, x_{N_{0}-1} ) \}, \nonumber
\end{eqnarray}
where $\phi_{\lambda}(x_{N_{0}})$, $\phi_{\mu } (x_{N_{0} })$, $\cdot 
\cdot \cdot$ and $\phi_{\xi}(x_{N_{0}})$ 
are the constituent spin orbitals of 
$\Phi_{SSD} (x_{1} , \cdot \cdot \cdot ,x_{N_{0} } )$. 
In what follows, suppose that these spin orbitals are given as 
the solutions of simultaneous equations of previous work,\cite{2} 
and therefore they are orthonormal to each other.
$\Phi_{SSD}^i (x_{1}, \cdot \cdot \cdot, x_{N_{0}-1})$ 
$(1 \! \le \! i \! \le \! N_{0})$ 
in Eq. (\ref{eq3}) denote $(N_{0} - 1)$-electron SSDs 
that are defined as the minor determinants multiplied by 
$1/{\sqrt {(N_{0} - 1)!} }$. 
The PD that is calculated from Eq. (\ref{eq3}) is given by 
\begin{equation}
\label{eq4}
\gamma_{SSD}^{(2)} ({\rm {\bf r{r}'}};{\rm {\bf r{r}'}})_{N = N_{0} } = 
\frac{1}{N_{0}-2}\sum\limits_{i = 1}^{N_{0} } {\gamma _{SSD}^{(2)} 
({\rm {\bf r{r}'}};{\rm {\bf r{r}'}})_{N = N_{0} - 1}^i },
\end{equation}
where $\gamma_{SSD}^{(2)} 
({\rm {\bf r{r}'}};{\rm {\bf r{r}'}})_{N = N_{0}-1}^i$ 
is the PD calculated from 
$\Phi _{SSD}^i (x_{1} , \cdot \cdot \cdot ,x_{N_{0} - 1} )$. 

Likewise, the cofactor expansion of each 
$\Phi_{SSD}^i (x_{1} , \cdot \cdot \cdot ,x_{N_{0} - 1} )$ 
yields $(N_{0}-2)$-electron SSDs, which are denoted by 
$\Phi_{SSD}^{ij} (x_{1}, \cdot \cdot \cdot, x_{N_{0}-2})$ 
$(1 \! \le \! j \! \le \! N_{0} \!- \! 1)$. 
Then, each $\gamma_{SSD}^{(2)} 
({\rm {\bf r{r}'}};{\rm {\bf r{r}'}})_{N=N_{0} -1}^i$ 
is given by 
$\sum\limits_{j=1}^{N_{0}-1} {\gamma_{SSD}^{(2)} 
({\rm {\bf r{r}'}};{\rm {\bf r{r}'}})_{N = N_{0}-2}^{ij} }/(N_{0}-3)$, 
where 
$\gamma_{SSD}^{(2)} 
({\rm {\bf r{r}'}};{\rm {\bf r{r}'}})_{N = N_{0}-2}^{ij}$ 
is the PD calculated from 
$\Phi_{SSD}^{ij} (x_{1}, \cdot \cdot \cdot, x_{N_{0}-2})$. 
Thus, $\gamma_{SSD}^{(2)} ({\rm {\bf r{r}'}};{\rm {\bf r{r}'}})_{N = N_{0}}$ 
can be expressed by the sum of 
$\gamma_{SSD}^{(2)} ({\rm {\bf r{r}'}};{\rm {\bf r{r}'}})_{N=N_{0}-2}^{ij}$. 
By the repetition of this procedure, we arrive at 
\begin{equation}
\label{eq5}
\gamma_{SSD}^{(2)} ({\rm {\bf r{r}'}};{\rm {\bf r{r}'}})_{N = N_{0}} = 
\frac{1}{(N_{0} - 2)!}\sum\limits_{i = 1}^{N_{0} } {\sum\limits_{j = 
1}^{N_{0} - 1} { \cdot \cdot \cdot \sum\limits_{p = 1}^4 {\sum\limits_{q = 
1}^3 {\gamma _{SSD}^{(2)} ({\rm {\bf r{r}'}};{\rm {\bf r{r}'}})_{N = 2}^{ij 
\cdot \cdot \cdot pq} } } } } ,
\end{equation}
where $\gamma_{SSD}^{(2)} 
({\rm {\bf r{r}'}};{\rm {\bf r{r}'}})_{N = 2}^{ij \cdot \cdot \cdot pq}$ 
$(1 \! \le \! i \! \le \! N_{0},  
1 \! \le \! j \! \le \! N_{0}-1, \cdot \cdot \cdot, 
1 \! \le \! p \! \le \! 4,  1 \! \le \! q \! \le \!3)$ 
are PDs calculated from the two-electron SSDs. 
These two-electron SSDs, which are denoted by 
$\Phi_{SSD}^{ij \cdot \cdot \cdot pq} (x_{1} ,x_{2} )$, 
are obtained by the above-mentioned successive cofactor expansions. 

Multiplying both sides of Eq. (\ref{eq5}) by 
$\left| {f\left( {\left| {{\rm {\bf r}} - {\rm {\bf {r}'}}} \right|} \right)} 
\right|^{2}$, we finally get 
\begin{eqnarray}
\label{eq6}
\left| {f\left( {\left| {{\rm {\bf r}} - {\rm {\bf {r}'}}} \right|} \right)} 
\right|^{2}\gamma_{SSD}^{(2)} ({\rm {\bf r{r}'}};{\rm {\bf r{r}'}})_{N=N_{0}} 
\!=\! \frac{1}{(N_{0} - 2)!}\sum\limits_{i = 1}^{N_{0} } {\sum\limits_{j 
= 1}^{N_{0} - 1} {\!\!\cdot \cdot \cdot \!\! 
\sum\limits_{p = 1}^4 {\sum\limits_{q = 
1}^3 {\left| {f\left( {\left| {{\rm {\bf r}} - {\rm {\bf {r}'}}} \right|} 
\right)} \right|\,^{2}\gamma _{SSD}^{(2)} ({\rm {\bf r{r}'}};{\rm {\bf 
r{r}'}})_{N = 2}^{ij \cdot \cdot \cdot pq} } } } }.\nonumber \\
\end{eqnarray}
This relation is the starting point to examine the $N$-representability of 
the LO-Jastrow PD. 
\section {Sufficient conditions for the {\textit{N}}-representablity of 
the LO-Jastrow PD}
We shall start with considering the $N$-representability of 
$\left| {f\left({\left| {{\rm {\bf r}} - {\rm {\bf {r}'}}} \right|} \right)} 
\right|^{2}\gamma_{SSD}^{(2)} ({\rm {\bf r{r}'}};{\rm {\bf r{r}'}})_{N = 
2}^{ij \cdot \cdot \cdot pq} $ 
that appears in the right-hand side of Eq. (\ref{eq6}). 
Suppose that the two-electron wave function that yields 
$\left| {f\left( {\left| {{\rm {\bf r}} - {\rm {\bf {r}'}}} \right|} \right)} 
\right|^{2}\gamma_{SSD}^{(2)} ({\rm {\bf r{r}'}};{\rm {\bf r{r}'}})_{N = 
2}^{ij \cdot \cdot \cdot pq} $ 
is given by the following Jastrow wave function;
\begin{equation}
\label{eq7}
\Psi_{N = 2}^{ij \cdot \cdot \cdot pq} (x_{1} ,\,x_{2} ) = \frac{1}{\sqrt 
{A_2^{ij \cdot \cdot \cdot pq} } }f\left( {\left| {{\rm {\bf r}}_{1} - {\rm 
{\bf r}}_{2} } \right|} \right)\Phi _{SSD}^{ij \cdot \cdot \cdot pq} (x_{1} 
,x_{2} ),
\end{equation}
where $A_2^{ij \cdot \cdot \cdot pq} $ is the normalization constant. 
From Eq. (\ref{eq7}), the PD is calculated as 
$\left| {f\left( {\left| {{\rm {\bf r}} - {\rm {\bf {r}'}}} \right|} \right)} 
\right|^{2}\gamma_{SSD}^{(2)} ({\rm {\bf r{r}'}};{\rm {\bf r{r}'}})_{N = 2}
^{ij \cdot \cdot \cdot pq} / A_2^{ij \cdot \cdot \cdot pq} $. 
Therefore, we get $A_2^{ij \cdot \cdot \cdot pq} = 1$ 
as the sufficient condition for the $N$-representability of 
$\left| {f\left( {\left| {{\rm {\bf r}} - {\rm {\bf {r}'}}} \right|} \right)} 
\right|^{2}\gamma _{SSD}^{(2)} ({\rm {\bf r{r}'}};{\rm {\bf r{r}'}})
_{N = 2}^{ij \cdot \cdot \cdot pq} $. 
Hereafter, we assume that such conditions hold for all values of 
$i,j,\cdot \cdot \cdot, p$ and $q$, i.e., 
\begin{equation}
\label{eq8}
A_2^{ij \cdot \cdot \cdot pq} = 1
\,\,\,\,\,
\mbox{for all values of } i,j,\cdot \cdot \cdot, p \mbox{ and } q.
\end{equation}
Note that the normalization constant $A_2^{ij \cdot \cdot \cdot pq} $ is 
determined by both 
$f\left( {\left| {{\rm {\bf r}}_{1} - {\rm {\bf r}}_{2} } \right|} \right)$ 
and 
$\Phi_{SSD}^{ij \cdot \cdot \cdot pq} (x_{1} ,x_{2} )$.\cite{5,6} 
Therefore, if they are given, we can calculate $A_2^{ij \cdot \cdot \cdot pq}$, and check whether the conditions Eq. (\ref{eq8}) are satisfied or not. 
As mentioned later, the conditions Eq. (\ref{eq8}) are the parts 
of sufficient conditions for the $N$-representability of Eq. (\ref{eq2}). 

Under the conditions Eq. (\ref{eq8}), we have
\begin{equation}
\label{eq9}
\left| {f\left( {\left| {{\rm {\bf r}} - {\rm {\bf {r}'}}} \right|} \right)} 
\right|^{2}\gamma _{SSD}^{(2)} ({\rm {\bf r{r}'}};{\rm {\bf r{r}'}})_{N = 
2}^{ij \cdot \cdot \cdot pq} = \left\langle {\Psi _{N = 2}^{ij \cdot \cdot 
\cdot pq} } \right|\hat {\gamma }^{(2)}({\rm {\bf r{r}'}};{\rm {\bf 
r{r}'}})_{N = 2} \left| {\Psi _{N = 2}^{ij \cdot \cdot \cdot pq} } 
\right\rangle ,
\end{equation}
where $\hat {\gamma }^{(2)}({\rm {\bf r{r}'}};{\rm {\bf r{r}'}})_{N = n} $ 
denotes the PD operator for an $n$-electron system. 
Substituting Eq. (\ref{eq9}) into Eq. (\ref{eq6}) and rearranging, we get
\begin{eqnarray}
\label{eq10}
&&\left| {f\left( {\left| {{\rm {\bf r}} - {\rm {\bf {r}'}}} \right|} 
\right)} \right|^{2}\gamma_{SSD}^{(2)} ({\rm {\bf r{r}'}};{\rm {\bf 
r{r}'}})_{N = N_{0} } \nonumber \\ 
&&\!\!\! = \!\!
\sum\limits_{i=1}^{N_{0}} { \!\! \left[ \!
\frac{1}{N_{0}\! -\! 2}\!\!\sum\limits_{j=1}^{N_{0} - 1} { \!\! \left[ \!
\frac{1}{N_{0}\! -\! 3}\!\!\sum\limits_{k=1}^{N_{0} - 2} { \!\! \left[ \! 
\cdot \cdot \cdot 
\frac{1}{3}\!\sum\limits_{o=1}^5 {\! \left[ \! {
\frac{1}{2}\!\sum\limits_{p=1}^4 {\! \left[ \! {
\frac{1}{1}\!\sum\limits_{q=1}^3 {\! \left 
\langle {\Psi_{N=2}^{ijk \cdot \cdot \cdot opq} } \right|
\hat {\gamma }^{(\ref{eq2})}({\rm {\bf r{r}'}};{\rm {\bf r{r}'}})_{N = 2} 
\left| {\Psi_{N=2}^{ijk \cdot \cdot \cdot opq}} \right\rangle }} 
\right]} } \right]} \right] } \right] } \right] } \nonumber \\
\end{eqnarray}
It should be noticed that the right-hand side of Eq. (\ref{eq10}) has a 
characteristic form. Concerning the $N$-representability of this form, 
the following theorem holds:

\noindent
\textit{Theorem.} If there exists the set of functions 
$\left\{ {a_{\alpha } (x_{n + 1} )} \right\}$ 
$(1 \! \le \! \alpha \! \le \! n + 1)$ that satisfy the conditions;
\begin{equation}
\label{eq11}
\int {a_\alpha ^\ast (x_{n + 1}) a_{{\alpha }'} (x_{n + 1})
\mbox{d}x_{n + 1} } = \frac{1}{n + 1}\delta_{\alpha {\alpha }'} ,
\end{equation}
\begin{equation}
\label{eq12}
\sigma \sum\limits_{\alpha = 1}^{n + 1} {a_{\alpha } (x_{n + 1} )\Psi _{N = 
n}^\alpha (x_{1} , \cdot \cdot \cdot ,x_{n} )} = ( - 1)^{\bar {\sigma 
}}\sum\limits_{\alpha = 1}^{n + 1} {a_{\alpha } (x_{n + 1} )\Psi _{N = 
n}^\alpha (x_{1} , \cdot \cdot \cdot ,x_{n} )} ,
\end{equation}
then the following equations hold:
\begin{equation}
\label{eq13}
\frac{1}{n-1}\sum\limits_{\alpha=1}^{n+1} {\left\langle {\Psi_{N=n}^\alpha } 
\right|\hat {\gamma }^{(2)}({\rm {\bf r{r}'}};{\rm {\bf r{r}'}})_{N = n} 
\left| {\Psi_{N = n}^\alpha } \right\rangle } = 
\left\langle {\Psi_{N = n + 1} } \right|\hat {\gamma }^{(2)}({\rm {\bf 
r{r}'}};{\rm {\bf r{r}'}})_{N = n + 1} \left| {\Psi _{N = n + 1} } 
\right\rangle 
\end{equation}
with
\begin{equation}
\label{eq14}
\Psi_{N = n + 1} (x_{1} , \cdot \cdot \cdot ,x_{n + 1} ) = 
\sum\limits_{\alpha = 1}^{n + 1} {a_{\alpha } (x_{n + 1} )\Psi _{N = 
n}^\alpha (x_{1} , \cdot \cdot \cdot ,x_{n} )} .
\end{equation}
Here $\Psi_{N = n}^\alpha (x_{1} ,\, \cdot \cdot \cdot ,\,x_{n} )$ 
$(1 \! \le \! \alpha \! \le \! n + 1)$ 
denote the $n$-electron wave functions, and $\sigma $ is 
a permutation operator upon the electron coordinates, and $\bar {\sigma }$ 
is the number of interchanges in $\sigma $. 

\noindent
\textit{Proof.} The left-hand side of Eq. (\ref{eq13}) seems to be 
related to the average of 
$\hat {\gamma }^{(\ref{eq2})}({\rm {\bf r{r}'}};{\rm {\bf r{r}'}})_{N = n} $ 
with respect to a density matrix for a mixed state. Indeed, 
if the density matrix for the mixed state is given by
\begin{equation}
\label{eq15}
\hat {\rho }_{n} = \sum\limits_{\alpha = 1}^{n + 1} {\frac{1}{n + 1}\left| 
{\Psi _{N = n}^\alpha } \right\rangle \left\langle {\Psi _{N = n}^\alpha } 
\right|} ,
\end{equation}
then the average of $\hat {\gamma }^{(\ref{eq2})}({\rm {\bf r{r}'}};{\rm {\bf 
r{r}'}})_{N = n} $ is calculated as
\begin{equation}
\label{eq16}
\mbox{Tr}[\hat {\rho }_{n} \hat {\gamma }^{(2)}({\rm {\bf r{r}'}};{\rm {\bf 
r{r}'}})_{N = n} ] = \frac{1}{n + 1}\sum\limits_{\alpha = 1}^{n + 1} 
{\left\langle {\Psi _{N = n}^\alpha } \right|\hat {\gamma }^{(2)}({\rm {\bf 
r{r}'}};{\rm {\bf r{r}'}})_{N = n} \left| {\Psi _{N = n}^\alpha } 
\right\rangle } .
\end{equation}
By using Eq. (\ref{eq16}), the left-hand side of Eq. (\ref{eq13}) 
is rewritten as 
$ \{(n \!+\! 1)/(n \!-\! 1) \} 
\mbox{Tr}[\hat {\rho }_{n} \hat {\gamma }^{(2)}
({\rm {\bf r{r}'}};{\rm {\bf r{r}'}})_{N = n} ]$. 

On the other hand, it is expected that the average 
$\mbox{Tr}[\hat {\rho }_{n} \hat {\gamma }^{(2)}
({\rm {\bf r{r}'}};{\rm {\bf r{r}'}})_{N = n} ]$ 
may be given as the expectation value of 
$\hat {\gamma }^{(2)}({\rm {\bf r{r}'}};{\rm {\bf r{r}'}})_{N = n} $ 
with respect to a pure state for the 
whole system that includes the 
$n$-electron system as a subsystem.\cite{13,14} 
We shall take an $(n + 1)$-electron system as the whole system, and suppose 
that the wave function for such the $(n + 1)$-electron system is given by 
Eq. (\ref{eq14}) with Eq. (\ref{eq11}). Then we indeed get
\begin{equation}
\label{eq17}
\left\langle {\Psi _{n + 1} } \right|\hat {\gamma }^{(\ref{eq2})}({\rm {\bf 
r{r}'}};{\rm {\bf r{r}'}})_{N = n} \left| {\Psi _{n + 1} } \right\rangle = 
\mbox{Tr}[\hat {\rho }_{n} \hat {\gamma }^{(\ref{eq2})}({\rm {\bf r{r}'}};{\rm {\bf 
r{r}'}})_{N = n} ].
\end{equation}
Furthermore, if $\Psi _{N = n + 1} (x_{1} , \cdot \cdot \cdot ,x_{n + 1} )$ 
is antisymmetric, i.e., if Eq. (\ref{eq12}) holds, 
then the expectation value of 
$\hat {\gamma }^{(\ref{eq2})}({\rm {\bf r{r}'}};{\rm {\bf r{r}'}})_{N=n+1}$ 
with respect to $\Psi_{N = n + 1} (x_{1} , \cdot \cdot \cdot ,x_{n + 1} )$ 
is given by 
\begin{equation}
\label{eq18}
\left\langle {\Psi _{n + 1} } \right|\hat {\gamma }^{(2)}({\rm {\bf 
r{r}'}};{\rm {\bf r{r}'}})_{N = n + 1} \left| {\Psi _{n + 1} } \right\rangle 
= \frac{n + 1}{n - 1}\left\langle {\Psi _{n + 1} } \right|\hat {\gamma}^{(2)}
({\rm {\bf r{r}'}};{\rm {\bf r{r}'}})_{N = n} \left| {\Psi _{n + 1} } 
\right\rangle .
\end{equation}
Consequently, Eq. (\ref{eq13}) immediately follows from 
Eqs. (\ref{eq16}), (\ref{eq17}) and (\ref{eq18}). 
This means that Eq. (\ref{eq13}) holds 
under the conditions that there exists the set of functions 
$\left\{ {a_{\alpha } (x_{n + 1} )} \right\}$ 
$(1 \! \le \! \alpha \! \le \! n \! + \! 1)$ 
that satisfy Eqs. (\ref{eq11}) and (\ref{eq12}). \,\,\,  Q.E.D.

First, we apply the above theorem to the term 
$\frac{1}{1}\sum\limits_{q = 1}^3 
{\left\langle {\Psi _{N = 2}^{ijk \cdot \cdot \cdot opq} } \right|
\hat {\gamma }^{(2)}({\rm {\bf r{r}'}};{\rm {\bf r{r}'}})_{N = 2} 
\left| {\Psi _{N = 2}^{ijk \cdot \cdot \cdot opq} } \right\rangle } $ 
that appears in Eq. (\ref{eq10}). 
According to the theorem, this term can be rewritten as
\begin{equation}
\label{eq19}
\frac{1}{1}\sum\limits_{q = 1}^3 {\left\langle {\Psi _{N = 2}^{ijk \cdot 
\cdot \cdot opq} } \right|\hat {\gamma }^{(2)}({\rm {\bf r{r}'}};{\rm {\bf 
r{r}'}})_{N = 2} \left| {\Psi _{N = 2}^{ijk \cdot \cdot \cdot opq} } 
\right\rangle } = \left\langle {\Psi _{N = 3}^{ijk \cdot \cdot \cdot op} } 
\right|\hat {\gamma }^{(2)}({\rm {\bf r{r}'}};{\rm {\bf r{r}'}})_{N = 3} 
\left| {\Psi _{N = 3}^{ijk \cdot \cdot \cdot op} } \right\rangle 
\end{equation}
with
\begin{equation}
\label{eq20}
\Psi_{N = 3}^{ijk \cdot \cdot \cdot op} (x_{1} ,\,x_{2} ,x_{3} ) = 
\sum\limits_{q = 1}^3 {a_q^{ijk \cdot \cdot \cdot op} (x_{3} )\Psi _{N = 
2}^{ijk \cdot \cdot \cdot opq} (x_{1} ,\,x_{2} )} ,
\end{equation}
if there exists the set of functions $\left\{ {a_q^{ijk \cdot \cdot \cdot 
op} (x_{3} )} \right\}$ $(1 \! \le \! q \! \le \! 3)$ 
that satisfy the following conditions:
\begin{equation}
\label{eq21}
\int {a_q^{ijk \cdot \cdot \cdot op} (x_{3} )^{\ast }a_{q'}^{ijk \cdot \cdot 
\cdot op} (x_{3} )\mbox{d}x = } \frac{1}{3}\delta _{q,{q}'} ,
\end{equation}
\begin{equation}
\label{eq22}
\sigma \sum\limits_{q = 1}^3 {a_q^{ijk \cdot \cdot \cdot op} (x_{3} )\Psi 
_{N = 2}^{ijk \cdot \cdot \cdot opq} (x_{1} ,x_{2} )} = ( - 1)^{\bar {\sigma 
}}\sum\limits_{q = 1}^3 {a_q^{ijk \cdot \cdot \cdot op} (x_{3} )\Psi _{N = 
2}^{ijk \cdot \cdot \cdot opq} (x_{1} ,x_{2} )} .
\end{equation}
In addition to Eq. (\ref{eq8}), the existence conditions for 
$\left\{ {a_q^{ijk \cdot \cdot \cdot op} (x_{3})} \right\}$ 
$(1 \! \le \! q \! \le \!3)$, 
i.e., Eqs. (\ref{eq21}) and (\ref{eq22}), are also the parts of sufficient 
conditions for the $N$-representability of Eq. (\ref{eq2}). 
We assume that the set of functions 
$\left\{ {a_q^{ijk \cdot \cdot \cdot op} (x_{3})} \right\}$ 
$(1 \! \le \! q \! \le \!3)$ 
is obtained. Substitution of Eq. (\ref{eq20}) into Eq. (\ref{eq10}) leads to
\begin{eqnarray}
\label{eq23}
&&\left| {f\left( {\left| {{\rm {\bf r}} - {\rm {\bf {r}'}}} \right|} 
\right)} \right|^{2}\gamma_{SSD}^{(2)} ({\rm {\bf r{r}'}};{\rm {\bf 
r{r}'}})_{N = N_{0} } \nonumber \\ 
&&\!\!\! = \!\!
\sum\limits_{i=1}^{N_{0}} { \!\! \left[ \!
\frac{1}{N_{0}\! -\! 2}\!\!\sum\limits_{j=1}^{N_{0} - 1} { \!\! \left[ \!
\frac{1}{N_{0}\! -\! 3}\!\!\sum\limits_{k=1}^{N_{0} - 2} { \!\! \left[ \! 
\cdot \cdot \cdot 
\frac{1}{3}\!\sum\limits_{o=1}^5 {\! \left[ \! {
\frac{1}{2}\!\sum\limits_{p=1}^4 {\! \! {
\left\langle {\Psi _{N = 3}^{ijk \cdot \cdot \cdot op} } \right|\hat 
{\gamma }^{(2)}({\rm {\bf r{r}'}};{\rm {\bf r{r}'}})_{N = 3} \left| 
{\Psi _{N = 3}^{ijk \cdot \cdot \cdot op} } \right\rangle } 
} } \right]} \right] } \right] } \right] }. 
\end{eqnarray}

Next, we apply the theorem to the term 
$\frac{1}{2}\sum\limits_{p = 1}^4 {\left\langle {
\Psi _{N = 3}^{ijk \cdot \cdot \cdot op} } \right|\hat 
{\gamma }^{(2)}({\rm {\bf r{r}'}};{\rm {\bf r{r}'}})_{N = 3} \left| {
\Psi _{N = 3}^{ijk \cdot \cdot \cdot op} } \right\rangle } $ 
that appears in Eq. (\ref{eq23}).  Similarly to the above, we obtain 
\begin{equation}
\label{eq24}
\frac{1}{2}\sum\limits_{p = 1}^4 {\left\langle {\Psi _{N = 3}^{ijk \cdot 
\cdot \cdot op} } \right|\hat {\gamma }^{(2)}({\rm {\bf r{r}'}};{\rm {\bf 
r{r}'}})_{N = 3} \left| {\Psi _{N = 3}^{ijk \cdot \cdot \cdot op} } 
\right\rangle } = \left\langle {\Psi _{N = 4}^{ijk \cdot \cdot \cdot o} } 
\right|\hat {\gamma }^{(2)}({\rm {\bf r{r}'}};{\rm {\bf r{r}'}})_{N = 4} 
\left| {\Psi _{N = 4}^{ijk \cdot \cdot \cdot o} } \right\rangle 
\end{equation}
with
\begin{equation}
\label{eq25}
\Psi _{N = 4}^{ijk \cdot \cdot \cdot o} (x_{1} , \cdot \cdot \cdot ,x_{4} ) 
= \sum\limits_{p = 1}^4 {a_p^{ijk \cdot \cdot \cdot o} (x_{4} )\Psi _{N = 
3}^{ijk \cdot \cdot \cdot op} (x_{1} ,\,x_{2} ,\,x_{3} )} ,
\end{equation}
if there exists the set of functions 
$\{a_p^{ijk \cdot \cdot \cdot o} (x_{4} )\}$ 
$(1 \! \le \! p \! \le \! 4)$ that satisfy the following conditions:
\begin{equation}
\label{eq26}
\int {a_p^{ijk \cdot \cdot \cdot o} (x_{4} )^{\ast }a_{p'}^{ijk \cdot \cdot 
\cdot o} (x_{4} )\mbox{d}x = } \frac{1}{4}\delta _{p,{p}'} ,
\end{equation}
\begin{equation}
\label{eq27}
\sigma \sum\limits_{p = 1}^4 {a_p^{ijk \cdot \cdot \cdot o} (x_{4} )\Psi _{N 
= 3}^{ijk \cdot \cdot \cdot op} (x_{1} ,x_{2} ,\,x_{3} )} = ( - 1)^{\bar 
{\sigma }}\sum\limits_{p = 1}^4 {a_p^{ijk \cdot \cdot \cdot o} (x_{4} )\Psi 
_{N = 3}^{ijk \cdot \cdot \cdot op} (x_{1} ,x_{2} ,\,x_{3} )} .
\end{equation}
The existence conditions for $\{a_p^{ijk \cdot \cdot \cdot o} (x_{4} )\}$ 
$(1 \! \le \! p \! \le \! 4)$, i.e., Eqs. (\ref{eq26}) and (\ref{eq27}), 
are added to the set of 
sufficient conditions for the $N$-representability of Eq. (\ref{eq2}). At this stage, 
Eq. (\ref{eq8}) and the existence conditions for 
$\left\{ {a_q^{ijk \cdot \cdot \cdot op} (x_{3} )} \right\}$ 
$(1 \! \le \! q \! \le \!3)$ and 
$\{a_p^{ijk \cdot \cdot \cdot o} (x_{4} )\}$ 
$(1 \! \le \! p \! \le \! 4)$ belong to the set of sufficient conditions. 

Thus the theorem is applied repeatedly, so that further conditions are added 
to the set of sufficient conditions. Continuing until the conditions for 
$\{a_{i} (x_{N_{0} } )\}$ $(1 \! \le \! i \! \le \! N_{0})$, 
we eventually obtain 
sufficient conditions for the $N$-representability of Eq. (\ref{eq2}). 

\section{Concrete steps for constructing the {\textit{N}}-representable 
LO-Jastrow PD}
In the preceding section, the sufficient conditions for the 
$N$-representability of the LO-Jastrow PD are derived. In this section, we 
consider the concrete steps for searching the correlation function that 
meets these conditions or checking its existence. 

\begin{enumerate}
\item \label{s1}
First, we give a trial form of the correlation function. Using this, 
simultaneous equations for the $N_{0} $-electron system are solved in a 
self-consistent way.\cite{2} 

\item 
\label{s2}
Let us consider the SSD that consists of the resultant spin orbitals 
for the simultaneous equations. The SSD can generally be expanded using the 
cofactor. The SSD for the $N_{0} $-electron system is expanded along the 
$N_{0} $th row, then we get the $N_{0} $ number of SSDs for the $(N_{0} - 
1)$-electron system, i.e., 
$\Phi_{SSD}^i \left( {x_{1}, \cdot \cdot \cdot, x_{N_{0}-1}} \right)$ 
$(1 \! \le \! i \! \le \! N_{0} )$. Successively, 
each of the SSDs for the $(N_{0} - 1)$-electron system is expanded along the 
$(N_{0} - 1)$th row, and then the $(N_{0} - 1)$ number of SSDs for the 
$(N_{0} - 2)$-electron system, i.e., 
$\Phi_{SSD}^{ij} \left({x_{1}, \cdot \cdot \cdot, x_{N_{0}-2}} \right)$ 
$(1 \! \le \! j \! \le \! N_{0} \!-\! 1)$, 
can be obtained for each $i$. After that and later, the cofactor expansions 
are likewise repeated, and we finally arrive at the SSDs for the 
two-electron system, i.e.,$\Phi _{SSD}^{ijk \cdot \cdot \cdot \cdot opq} 
\left( {x_{1} ,\,x_{2} } \right)$. The number of two-electron SSDs thus 
obtained is ${N_{0} !} \mathord{\left/ {\vphantom {{N_{0} !} 2}} \right. 
\kern-\nulldelimiterspace} 2$, since $i,\,j,\,k, \cdot \cdot \cdot \cdot 
,\,o,\,p$ and $q$ are integers such that 
$1 \! \le \! i \! \le \! N_{0} $, 
$1 \! \le \! j \! \le \! N_{0}\! -\! 1$, 
$1 \! \le \! k \! \le \! N_{0}\! -\! 2$,$ \cdot \cdot \cdot \cdot $, 
$1 \! \le \! o \! \le \! 5$, 
$1 \! \le \! p \! \le \! 4$ and 
$1 \! \le \! q \! \le \! 3$, respectively. 

\item 
\label{s3}
The two-electron SSD $\Phi_{SSD}^{ijk \cdot \cdot \cdot \cdot opq} 
\left( {x_{1} ,\,x_{2} } \right)$ and the correlation function determine the 
normalization constant $A_2^{ijk \cdot \cdot \cdot \cdot opq}$.\cite{5,6} 
We check whether all of $A_2^{ijk \cdot \cdot \cdot \cdot opq}$ are unity or 
not. If no, we return to the step \ref{s1} and change 
the form of the correlation function. This process proceeds 
until all of $A_2^{ijk \cdot \cdot \cdot \cdot opq} $ become unity. 
Suppose that such the correlation function is 
found, we get two-electron antisymmetric wave functions $\Psi _{N = 2}^{ijk 
\cdot \cdot \cdot \cdot opq} \left( {x_{1} ,\,x_{2} } \right)$ for all cases 
of $i,\,j,\,k,\, \cdot \cdot \cdot \cdot ,\,o,\,p$ and $q$ 
through Eq. (\ref{eq7}).

\item
\label{s4}
By means of these $\Psi_{N = 2}^{ijk \cdot \cdot \cdot \cdot opq} \left( 
{x_{1} ,\,x_{2} } \right)$, three-electron wave functions are defined as Eq. 
(\ref{eq20}). We will check whether there exists the set of 
$\left\{ {a_q^{ijk \cdot \cdot \cdot \cdot op} (x_{3} ),\,\,q = 1,\,2,\,3} 
\right\}$ that are satisfied with Eqs. (\ref{eq21}) and (\ref{eq22}). 
Note that the check has to be performed for all cases of 
$i,\,j,\,k,\, \cdot \cdot \cdot \cdot ,\,o$ and $p$. 
If suitable 
$\left\{ {a_q^{ijk \cdot \cdot \cdot \cdot op} (x_{3} )} \right\}$ 
cannot be found, we again return to the step \ref{s1} 
and modify the correlation function. Suppose the suitable 
$\left\{ {a_q^{ijk \cdot \cdot \cdot \cdot op} (x_{3} )} \right\}$ 
is found in this process, then three-electron antisymmetric wave functions 
$\Psi_{N = 3}^{ijk \cdot \cdot \cdot \cdot op} 
\left( {x_{1} ,\,x_{2} ,x_{3} } \right)$ 
for all cases of $i,\,j,\,k,\, \cdot \cdot \cdot \cdot ,\,o$ and $p$ 
can be constructed from Eq. (\ref{eq20}). 

\item
\label{s5}
We successively proceed the case of four-electron wave functions that are 
defined as Eq. (\ref{eq25}). In a similar way to the step \ref{s4}, 
we check whether the set of 
$\left\{ {a_p^{ijk \cdot \cdot \cdot \cdot o} (x_{4} ),\,\,p = 
1,\,2,\,3,\,4} \right\}$ 
meets the conditions of Eqs. (\ref{eq26}) and (\ref{eq27}). If no, 
we restart from the step \ref{s1} with the modified correlation function. 
Suppose that the suitable 
$\left\{ {a_p^{ijk \cdot \cdot \cdot \cdot o} (x_{4} )} \right\}$ 
are found for all cases of $i,\,j,\,k,\, \cdot \cdot \cdot \cdot $ and $o$, 
the four-electron antisymmetric wave functions 
$\Psi _{N = 4}^{ijk \cdot \cdot \cdot \cdot o} 
\left( {x_{1} ,\, \cdot \cdot ,\,x_{4} } \right)$ 
can be obtained from Eq. (\ref{eq25}).

\item
\label{s6}
Likewise, we further proceed the problem constructing the antisymmetric wave 
functions for more-electron systems. We search the set of 
$\left\{ {a_{\alpha } (x_{n + 1} ),\,\,
\alpha = 1,\,2,\, \cdot \cdot \cdot \cdot ,\,n + 1} \right\}$ 
that are satisfied with Eqs. (\ref{eq11}) and (\ref{eq12}), together with 
modifying the correlation function. If we successfully find the correlation 
function that meets the conditions (\ref{eq11}) and (\ref{eq12}) for any 
$n\,\,( \le N_{0} - 1)$, the LO-Jastrow PD of 
the $N_{0}$-electron system becomes $N$-representable. 
\end{enumerate}

As easily inferred, the above steps are feasible only for the small-electron 
systems from the practical viewpoint. However, it should be noted that the 
correlation function that makes the LO-Jastrow PD $N$-representable may, in 
principle, be found along the above steps, though there is a possibility 
that the suitable correlation function may not exist in some system. 

\section{Concluding remarks}
In this paper, the sufficient conditions for the $N$-representability of the 
LO-Jastrow PD are discussed. Using the properties of the LO-Jastrow PD, we 
derive the sufficient conditions that are imposed on the correlation 
function of the Jastrow wave function. As shown in Sec. IV, additional steps 
to search the suitable correlation function, which satisfies the sufficient 
conditions, are attached to the computational scheme proposed 
previously.\cite{2} 
Although the number of steps rapidly increases with that of electrons, 
the concrete steps that are presented in Sec. IV are feasible for a 
small-electron system. Of course, there is a possibility that the suitable 
correlation function cannot be found out. In this case, as mentioned in the 
previous paper, LO-Jastrow PDs are approximately $N$-representable in a 
sense that they satisfy 
four kinds of necessary conditions.\cite{2,7,8,9,10,11,12}

\begin{acknowledgments}
This work was partially supported by Grant-in-Aid for Scientific Research 
(No. 19540399) and for Scientific Research in Priority Areas (No. 17064006) 
of The Ministry of Education, Culture, Sports, Science, and Technology, 
Japan.
\end{acknowledgments}


\end{document}